\documentclass[american,aps,pra,reprint,groupedaddress,twocolumn,showpacs]{revtex4-1} 
\usepackage[unicode=true,pdfusetitle, bookmarks=true,bookmarksnumbered=false,bookmarksopen=false, breaklinks=false,pdfborder={0 0 0},backref=false,colorlinks=false] {hyperref}
\hypersetup{ colorlinks,linkcolor=myurlcolor,citecolor=myurlcolor,urlcolor=myurlcolor}
\usepackage{braket,colortbl,amsthm,amsmath,amssymb,txfonts,cleveref}
\definecolor{myurlcolor}{rgb}{0,0,0.9}
\usepackage{graphics,graphicx}
\usepackage{color}
\usepackage{colortbl}
\usepackage{subfigure}
\usepackage{graphicx}
\usepackage{ragged2e}
\usepackage{amsmath}
\usepackage{graphicx}
\usepackage{color}
\usepackage{amsmath}
\usepackage{tikz}
\usetikzlibrary{shapes.geometric, arrows}
\usepackage{latexsym}
\usepackage{amssymb}
\usepackage{amsthm}
\usepackage{bm}
\usepackage{graphics,epstopdf}
\usepackage{color}\usepackage{amsmath}

\usepackage[usenames,dvipsnames,svgnames]{pstricks}
\usepackage{epsfig}
\usepackage{pst-grad} 
\usepackage{pst-plot} 
\tikzstyle{startstop} = [rectangle, rounded corners, minimum width=3cm, minimum height=1cm,text centered, draw=black, fill=red!30]

\tikzstyle{env}=[circle,  ball color = green!20, minimum size= 80mm]
\tikzstyle{central}=[circle, ball color = red!100, minimum size=8mm]
\tikzstyle{bath}=[circle, ball color =blue!75, minimum size=4mm]

\theoremstyle{plain}

\def\bea{\begin{eqnarray}}
\def\eea{\end{eqnarray}}
\def\ba{\begin{array}}
\def\ea{\end{array}}

\def\beq{\begin{equation}}
\def\eeq{\end{equation}}

 \DeclareMathOperator\arctanh{arctanh}

\begin{document}
\title{Quantum precision thermometry with weak measurement}

\author{Arun Kumar Pati}
\email{akpati@hri.res.in}
\affiliation{Quantum Information and Computation Group, Harish-Chandra Research Institute, HBNI, Allahabad 211019,  India}
\author{Chiranjib Mukhopadhyay}
\email{chiranjibmukhopadhyay@hri.res.in}
\affiliation{Quantum Information and Computation Group, Harish-Chandra Research Institute, HBNI, Allahabad 211019,  India}
\author{Sagnik Chakraborty}
\email{csagnik@imsc.res.in}
\affiliation{Optics \& Quantum Information Group, The Institute of Mathematical Sciences, HBNI, C. I. T. Campus, Taramani, Chennai - 600 113, India}
\author{Sibasish Ghosh}
\email{sibasish@imsc.res.in}
\affiliation{Optics \& Quantum Information Group, The Institute of Mathematical Sciences, HBNI, C. I. T. Campus, Taramani, Chennai - 600 113, India}

\begin{abstract}
{\small{As the minituarization of electronic devices, which are sensitive to temperature,  grows apace, sensing of temperature with ever smaller probes is more important than ever.  Genuinely quantum mechanical schemes of thermometry are thus expected to be crucial to future technological progress. We propose a new method to measure the temperature of a bath using the weak measurement scheme with a finite dimensional probe. The precision offered by the present scheme 
not only shows similar qualitative features as the usual Quantum Fisher Information based thermometric protocols, but also allows for flexibility over setting the optimal thermometric window through judicious choice of post selection measurements.
}}   
\end{abstract}

\maketitle
\section{Introduction}

Quantum mechanics lends different status to observable quantities and parameters of a system, the latter of which are not expressible through Hermitian operators. Among physical parameters, temperature of a system is perhaps one of the most ubiquitously useful. Temperature drives chemical reaction rates \citep{arrhenius}, control the generation of thermomelectric current in thermocouples \citep{seebeck}, or heat flow between different baths \citep{callenbook}.  Thus, thermometry, or the science of measuring temperature, is of paramount importance. This is especially relevant for modern technological applications and devices, where the thermal baths may themselves be relatively tiny. Nanoscale probes, which follow laws of quantum physics, are thus required to measure the temperature of such baths without disturbing them too much. Quantum thermometry thus aims at  improving the technique of temperature sensing using small probes \citep{hofer, ivanov, thermo_review, PhysRevA.97.032129, sanpera_review, qt1, qt2}. The analysis of quantum thermometry so far has focused on seeking to find and saturate the \emph{quantum Cramer Rao bound} for various partially and fully thermalized configurations of systems \citep{correa_prl, barbieri_pra_recent}. In addition to ensure non-invasiveness, these studies show possible improvement in the precision due to quantum effects for non-equilbrium settings. Even in the steady state scenario, quantum enhancement in the precision using a quantum switch has been recently reported \citep{our_thermometry}. For observables, another technique to experimentally determine them quantitatively has recently come to the fore, which relies on a so called $weak$ $measurement$ scheme \citep{AAV, FXS, HBL, JMH}. Weak measurement is a technique of ascertaining information about an observable through a \emph{weak} interaction between the system and the measurement apparatus generated by the observable. It is then followed by a strong \emph{post-selection} measurement on the system. One of the most well known facets of weak measurement scheme is the concept of the so called \emph{weak values} of an observable, which are, in general, complex numbers. In the last few years, weak values have been used in the context of measuring quantities which may or may not have quantum mechanical observables associated with them, for example, the geometric phase \citep{weakgeometricphase}, non Hermitian operators \citep{weaknonHermitian}, density matrix corresponding to a quantum state \citep{weakquantumstate1,weakquantumstate2,weakquantumstate3, jordan1, jordan2}, or the entanglement content of a quantum state \citep{weakentanglement1,weakentanglement2}. The weak value amplification technique has found recent physical application in observations of the spin Hall effect \citep{weakspinhall}, photon trajectories \citep{weakphotontrajectory}, or the time delay between ultra fast processes \citep{weaktimedelay} as well. Thus, it is reasonable to ask whether a weak measurement based scheme is a viable approach towards thermometry. In this paper,  we provide such a scheme. We mention here that there has been another theoretical as well as experimental paper by Egan and Stone \citep{egan1} \footnote{We thank an anonymous referee for bringing this work to our attention.} in the recent past, which introduces the concept of weak thermometry. Experiments demonstrating weak value enhanced metrological tasks have also ben performed in analogus contexts \citep{egan2, egan3}.

In the present work, we outline how to measure temperature using weak values with finite dimensional probes and specifically concentrate on qubit probes. We show the presence of an optimal temperature window for the precision offered by this scheme, which is a feature repeated in the usual paradigm of quantum thermometry \citep{stace, correa_prl, Xie2017, paris_therm}. One crucial advantage offered by the present scheme over previously considered schemes is the ability to shift the optimal precision window while keeping the probe parameters fixed, and simply changing the post selected state. In addition, we argue that, the proposal based on the weak value is suitable for measurement of a very hot body. If the temperature of a system is very high, then prolonged contact with a measuring apparatus may damage the apparatus itself. On the other hand, in our weak measurement based scheme, the measuring apparatus  is brought into contact with a thermalized qubit of the bath for a very short duration of time, thus potentially saving any damage to the apparatus. \footnote{According to the formulation of the weak measurement protocol, described in section \ref{sec3}, it is enough to consider $g \tau \ll 1$, even if we allow hot probes. Note that, for performing the post-selection strong measurement, the assumption $g \tau \ll 1$ may be dropped. Nevertheless, the issue of the measuring apparatus getting destroyed due to the hot probe can be circumvented by considering a different measuring apparatus. }

The organization of the paper is as follows. Section \ref{sec2} briefly reviews the weak measurement scheme. The main theme of our paper, namely the protocol for measuring temperature through weak values, is described in Section \ref{sec3}. This is followed by the detailed analysis of precision offered by the present scheme for a qubit probe in Section \ref{sec4}. Section \ref{sec5} contains a complementary QFI based analysis of precision for the weak thermometric protocol. We finally conclude with a few observations and outline some possible future developments. 

\section{Weak measurement}
\label{sec2}
In theory of weak measurements, a quantum system $S$ is chosen to be in an initial state $\ket{\psi_i}_S$, called the \emph{pre-selected} state, and it is made to interact with a measurement apparatus $M$, prepared in a state $\ket{\phi}_M$. The interaction is weak in strength, and generated by a Hamiltonian $H_{int}=g\hat{A}\otimes \hat{P}_x$, where $\hat{A}$ is a system observable, $\hat{P}_x$ is the momentum operator of $M$ and $g$ is a small positive number signifying the strength of the interaction. This is followed by a strong measurement on the system, called the \emph{post-selection} measurement. The outcome of post-selection measurement that we focus on is along the state $\ket{\psi_f}$. Clearly, the post-selection is a selective measurement, as we ignore the other outcomes of the measurement.


The time-evolved state of $S + M$ before post-selection  is given by
\begin{align}
 \ket{\Psi}&=e^{- ig {\hat{A}} \otimes {\hat{P}}_x}\Big[\ket{\psi_i}\otimes\ket{\phi}_M\Big]\nonumber\\
 &\approx  \Big(I_S \otimes I_M - ig {\hat{A}} \otimes {\hat{P}}_x\Big)\Big[\ket{\psi_i} \otimes \ket{\phi}_M\Big],
\end{align}
up to first order in $g$. Note that the state $\ket{{\tilde{\Psi}}}_{SM} \equiv (I_S \otimes I_M - ig {\hat{A}} \otimes {\hat{P}}_x)\big[\ket{{\psi}_i} \otimes \ket{\phi}_M\big]$ is, in general, unnormalized. The reduced  state of $\ket{\tilde{\Psi}}_{SM}$ on $M$ is
\begin{equation}
\ket{{\tilde{\phi}_f}}_M \approx \Big(\braket{\psi_f|\psi_i} - ig\bra{{\psi}_f}{\hat{A}}\ket{\psi}_i{\hat{P}}_x\Big)\ket{\phi}_M,
\end{equation} 
which on normalizing looks like
\begin{equation}
 \ket{\phi_f}_M \approx \Big(1 - ig A_w \hat{P}_x\Big)\ket{\phi}_M \approx e^{- ig A_w\hat{P}_x}\ket{\phi}_M,
\end{equation}
where $A_w \equiv \bra{{\psi}_f}\hat{A}\ket{\psi_i}/\braket{\psi_f|\psi_i}$ is called as the {\it weak value} of $\hat{A}$.  Note that $A_w$ can also be complex and take values beyond the eigenvalue range of the observable.

Since the weak value $A_w$ is, in general a complex quantity, one must experimentally observe the real as well as imaginary parts of the weak value. On measuring certain properties of $\ket{\phi_f}_M$, the real and imaginary parts of $A_w$ can be determined. These properties include shift in position and momentum values compared to that of $\ket{\phi}_M$, variance of momentum wave-function, rate of change of position wave-function and the strength of interaction \citep{Jozsa_weak}.  Laguerre-Gaussian modes of an optical beam have also been used to measure the real and imaginary parts of a weak value \citep{kobayashi, lgjordan}.  Recently it has also been proposed and experimentally demonstrated that the weak value can be inferred from interference visibility and phase shifts \citep{nirala}.

\begin{figure}
\includegraphics[width = 0.6 \linewidth]{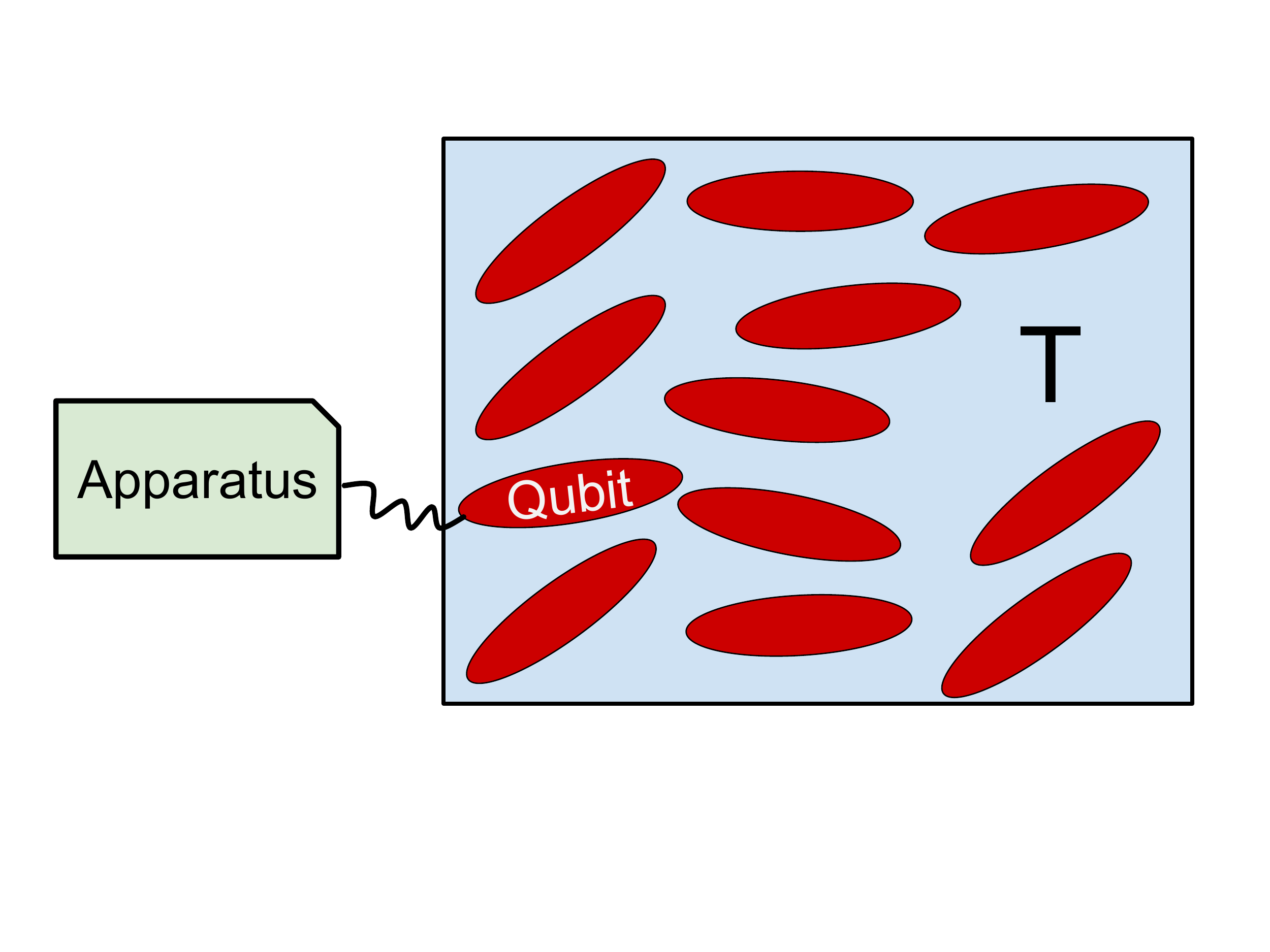}
\caption{(color online) Schematic of the thermometric scheme. We assume the bath being a collection of thermalized qubits, and couple the measuring apparatus to one of the qubits, which is called the probe qubit for a very short time.}
\end{figure}
        
\section{Assessing temperature through weak values}
\label{sec3}
Consider a $d$ dimensional quantum system $S$ under the action of a time-independent Hamiltonian $\hat{H}$ having non-degenerate energy eigenvalues $E_1, E_2, \dots, E_d$ and the corresponding energy eigenstates $\ket{\psi_1}, \ket{\psi_2}, \ldots, \ket{\psi_d}$. Assume that $S$ is in contact with a heat bath of temperature $T$ and $S$ has reached the thermal equilibrium state ${\rho}_T \equiv e^{- {\beta}\hat{H}}/~ ({\rm Tr} [e^{- {\beta}\hat{H}}]) = \Big(\sum_{n = 1}^{d} e^{- {\beta}E_n}\ket{\psi_n}\bra{\psi_n}\Big)/\Big(\sum_{n = 1}^{d} e^{- {\beta}E_n}\Big)$. Here $\beta = 1/(k_BT)$, with $k_B$ being taken as unity. 
Prepare the measuring aparatus $M$ in a state $|{\phi}\rangle$, having position wave function ${\phi}(x)$. Note that here $M$ is considered to be a continuous-variable system, in general. We would like to perform measurement of an observable $\hat{A}$ on  the system $S$.

Consider now evolution of $S + M$ under the action of an interaction Hamiltonian $\hat{H}_{int} = g \hat{A} \otimes \hat{P}_x$ for a small amount of time $\tau$. The interaction strength $g$ is also considered to be small -- in the regime of weak interaction between $S$ and $M$. Here $\hat{P}_x$ is the momentum observable of $M$ canonically conjugate to the position observable $\hat{X}$. We assume here that during the time interval $[0, \tau]$, $S$ and $M$ are under the action of {\it only} the Hamiltonian $\hat{H}_{int}$. This may be fulfilled in different ways: (i) We may decouple the system $S$ at time $t = 0$ from the heat bath (after $S$ achieves the thermal equilibrium state ${\rho}_T$) and thereby switch on the interaction Hamiltonian $\hat{H}_{int}$ for the time duration $[0, \tau]$. (ii) On the other hand, we may think of assuming here that the the strength $g$ of interaction is much higher than that of the system Hamiltonian $\hat{H}$, so that due to action of $\hat{H}_{int}$ for a small time span $\tau$, it is enough to consider the change in states of $S$ under the action of $\hat{H}_{int}$ only. The time span $\tau$ should be small enough so that in spite of taking the strength $g$ of the interaction Hamiltonian $H_{int}$ being greater than the free Hamiltonian, $g \tau \ll 1 $. At the end of the action of the interaction Hamiltonian, the joint state of $S + M$ becomes:
\begin{align}
 \rho_{SM}({\tau}) &= e^{- ig{\tau} {\hat{A}} \otimes {\hat{P}}_x}({\rho}_T \otimes |\phi{\rangle}{\langle}{\phi}|)e^{ig{\tau} {\hat{A}} \otimes {\hat{P}}_x}\nonumber\\
 &\approx (I_S \otimes I_M - ig{\tau} {\hat{A}} \otimes {\hat{P}}_x)(\rho_T \otimes \ket{\phi}\bra{\phi})\nonumber\\
 &\times(I_S \otimes I_M + ig{\tau} {\hat{A}} \otimes {\hat{P}}_x)\nonumber\\
 &\approx \rho_T \otimes \ket{\phi}\bra{\phi} - ig\tau[\hat{A} \otimes \hat{P}_x, \rho_T \otimes \ket{\phi}\bra{\phi}].\label{eqn1}
\end{align}
Now we post-select the state of $S$ to be $|{\psi}_f\rangle$. Then $M$ will get collapsed into the following (unnormalized) state:
\begin{align}
 &\tilde{\eta}({\tau}) = \braket{\psi_f|\rho_T|\psi_f}\ket{\phi}\bra{\phi}\nonumber\\
 &- ig\tau \Big(\bra{\psi_f}\hat{A}\rho_T\ket{\psi_f}\hat{P}_x\ket{\phi}\bra{\phi} - \bra{\psi_f}\rho_T\hat{A}\ket{\psi_f}\bra{\phi}\bra{\phi}{\hat{P}}_x\Big)\nonumber\\
 &= \bra{\psi_f}\rho_T\ket{\psi_f} \Big[\ket{\phi}\bra{\phi}\nonumber\\
 &-ig\tau\Big(A_w{\hat{P}}_x\ket{\phi}\bra{\phi} - A_w^*\ket{\phi}\bra{\phi}\hat{P}_x\Big)\Big]\nonumber\\
 &\approx \bra{\psi_f}\rho_T\ket{\psi_f} \times \eta({\tau}),
\end{align}
with the corresponding normalized collapsed state of $M$ being given by:
\begin{equation}
\label{eqn2}
\eta(\tau) = e^{- ig{\tau}A_w{\hat{P}}_x}\ket{\phi}\bra{\phi}e^{ig\tau A_w\hat{P}_x}
\end{equation}
and the corresponding weak value is given by:
\begin{equation}
\label{eqn3}
A_w = \frac{\bra{\psi_f}\hat{A}\rho_T\ket{\psi_f}}{\bra{\psi_f}\rho_T\ket{\psi_f}}.
\end{equation}  
Using the value of $A_w$ together with {\it a priori} knowledge of $|{\psi}_f\rangle$, $\hat{A}$, and the energy eigen spectrum of the system Hamiltonian $\hat{H}$, one can, in principle, find out the value of the temperature $T$ -- with the help of eqn. (\ref{eqn3}). Let us note in passing that the operator $A$ must not commute with the relevant energy eigenbasis, else the weak value ceases to depend on the inverse temperature $\beta$, and thus, measuring the weak value furnishes no thermometric advantage.

\emph{High temperature regime-} Let us now consider the case where the bath temperature is high, that is, $\beta \rightarrow 0$.  Thus, we can replace $e^{-\beta} \approx 1 - \beta$. Now, assuming the spectral decomposition of $\hat{A}$ with eigenvalues $a_j$ and corresponding eigenstates $|a_j\rangle$ for $j = 1, 2, \ldots, d$, in conjunction with eqn. (\ref{eqn3}), helps us to obtain the following expression for the weak value.
\begin{align}
&A_w = \frac{\sum_{j, k = 1}^{d} a_je^{- {\beta}E_k}\braket{\psi_f|a_j}\braket{a_j|\psi_k}\braket{\psi_k|\psi_f}}{\sum_{l = 1}^{d} e^{- {\beta}E_l}|\braket{\psi_f|\psi_l}|^2}\nonumber\\
&\approx \frac{\bra{\psi_f}\hat{A}\ket{\psi_f} - \beta\bra{\psi_f}\hat{A}\hat{H}\ket{\psi_f}}{1 - \beta\bra{\psi_f}\hat{H}\ket{\psi_f}}\nonumber\\
&\approx \Big(\bra{\psi_f}\hat{A}\ket{\psi_f} - \beta\bra{\psi_f}\hat{A}\hat{H}\ket{\psi_f}\Big)\Big(1 + {\beta}\bra{\psi_f}\hat{H}\ket{\psi_f}\Big)\label{eqn4}\\
&\approx \bra{\psi_f}\hat{A}\ket{\psi_f} \nonumber\\
&+ \beta \big(\bra{\psi_f}\hat{A}\ket{\psi_f}\times \bra{\psi_f}\hat{H}\ket{\psi_f} - \bra{\psi_f}\hat{A}\hat{H}\ket{\psi_f}\Big)\label{eqn5}
\end{align}
Inverting this expression, in the high temperature limit, the inverse temperature is expressible in terms of the weak value of the observable $A$ as
\begin{equation}
\label{eqn6}
\beta \approx \frac{A_w - {\langle}{\psi}_f|\hat{A}|{\psi}_f{\rangle}}{{\langle}{\psi}_f|\hat{A}|{\psi}_f{\rangle} \times {\langle}{\psi}_f|\hat{H}|{\psi}_f{\rangle} - {\langle}{\psi}_f|\hat{A}\hat{H}|{\psi}_f{\rangle}}.
\end{equation}

Let us now anayze the right hand side of the above result in further detail. We denote the standard deviation of an observable $O$ as $\Delta O$.  According to Vaidman's formula \citep{vaidman}, $A|\psi_f\rangle = \langle A \rangle|\psi_f\rangle + \Delta A |\bar{\psi_f}\rangle$, which implies the following expression for the weak value 
\begin{equation}
A_{w} = \langle A \rangle + \Delta A \frac{\langle \bar{\psi_f} |\rho_{T}|\psi_f\rangle}{\langle \psi_f |\rho_T | \psi_f \rangle}.
\end{equation}
\noindent Here $|\bar{\psi} \rangle$ indicates that it is a state orthogonal to $|\psi\rangle$. This formula has been proved \citep{vaidman} in the following way - we can always decompose the state $|\psi\rangle$ as $|\psi\rangle = \alpha |\psi\rangle + \beta |\bar{\psi} \rangle$. Now, $\langle A \rangle = \langle \psi |A |\psi \rangle = \alpha$, and $\Delta A = \sqrt{ \langle A^2 \rangle - \alpha^2} =\sqrt{ \langle \psi | A^{\dagger} A |\psi \rangle - \alpha^2} = \sqrt{\alpha^2 + \beta^2 - \alpha^2} = \beta$. Similarly applying Vaidman's  formula for the density operator $\rho_{T}$ yields $\rho_T |\psi_f\rangle = \langle \rho_T \rangle + \Delta \rho_T |\bar{\bar{\psi_f}} \rangle$, where $|\bar{\bar{\psi_f}}\rangle$ is another state perpendicular to $|\psi_f\rangle$. Plugging in this expression to the earlier equation for weak value of $A$ yields the following expression for the inverse temperature 
\begin{equation}
\beta = - \frac{\Delta A \Delta \rho_T}{ \text{Cov} (A,H)} \frac{\langle \bar{\psi_f} | \bar{\bar{\psi_f}} \rangle}{\langle \psi_f |\rho_T |\psi_f\rangle}
\end{equation}

For a qubit state, the corresponding orthogonal state is unique. Hence, $|\bar{\psi_f}\rangle = |\bar{\bar{\psi_f}}\rangle$, which, when plugged in,  yields the following expression 

\begin{equation}
\beta = - \frac{\Delta A \Delta \rho_T}{\text{Cov} (A,H) \langle \psi_f | \rho_T | \psi_f \rangle}.
\end{equation}

\noindent The above equation may be of independent interest. Let us now note that the anomalous weak value $\delta A =  |\text{Re} (A_w) - \langle A \rangle|$, is expressible as $\delta A = |\text{Cov} (A,\rho_T)|/|\langle \rho_T \rangle|$. Now, $\langle \rho_T \rangle = \langle \exp(-\beta H) \rangle /Z \leq 1/\left( Z (1-\beta \langle H \rangle \right)$, where $Z$ is the corresponding canonical partition function. Combining these results together, we obtain the following lower bound for the temperature

\begin{equation}
T \geq \frac{\langle H \rangle}{1 - \frac{|\text{Cov}(A,\rho_T|}{Z \delta A}  }
\end{equation}

\noindent In the above equation, the anomalous weak value $\delta A$ is a truly quantum mechanical quantity. It is easy to note that the achievable lower bound on the temperature is stronger, if $\delta A$ is large in magnitude. 
 
\emph{Qubit case -} Let us now consider the generic situation in the case of the simplest non-trivial quantum system, which is a qubit, for arbitrary temperature. Consider $H=\sum_{i=1}^dE_i\ket{\psi_i}\bra{\psi_i}$. If we take now $\hat{A} = - i|{\psi}_1{\rangle}{\langle}{\psi}_2| + i|{\psi}_2{\rangle}{\langle}{\psi}_1|$ and $|{\psi}_f\rangle = (1/{\sqrt{2}})(|{\psi}_1\rangle + |{\psi}_2\rangle)$, then the weak value $A_w$ is given by:
\begin{equation}
A_w \equiv {\langle}{\psi}_f|\hat{A}{\rho}_T|{\psi}_f{\rangle}/{\langle}{\psi}_f|{\rho}_T|{\psi}_f{\rangle} = i  \frac{e^{- {\beta}E_1} - e^{- {\beta}E_2}}{e^{- {\beta}E_1} + e^{- {\beta}E_2}}.
\label{qubit_eqn}
\end{equation}
Note that here ${\langle}{\psi}_f|\hat{A}|{\psi}_f{\rangle} = 0$, ${\langle}{\psi}_f|\hat{H}|{\psi}_f{\rangle} = (E_1 + E_2)/2$, ${\langle}{\psi}_f|\hat{A}\hat{H}|{\psi}_f{\rangle} = i(E_1 - E_2)/2$. For the high temperature limit, we have the following (approximate) expression for the inverse temperature of the bath in the qubit case, which may be shown to be consistent with \eqref{eqn6}.
\begin{equation}
\label{eqn7}
\beta \approx \frac{- 2iA_w}{E_2 - E_1}.
\end{equation}

Now, based upon the existing methods of identifying/measuring the weak value $A_w$ \citep{Jozsa_weak, kobayashi}, one can, in principle, get an estimate of the (inverse) temperature $\beta$ in eqn. (\ref{eqn7}). One of the advantages of the present scheme is that the measuring apparatus is brought into contact with the heat bath for a very short time, hence reducing the chance of damage to the apparatus. We add here that the rationale behind choosing the operator $A$ in this form is the following, if $A$ is chosen as being along the z-axis of the Bloch sphere, there is no information to be obtained about the temperature from the weak value. In case of thermalization, the azimuthal symmetry of the state in the Bloch sphere picture means it is equally feasible choosing any operator $A$ along the $x-y$ plane of the Bloch sphere, thus we may assume the above form of $A$ without loss of generality. Once $A$ is fixed, the azimuthal symmetry of the problem is lost, and one has to choose the post selected state carefully. Above, we assumed an example of the post selected state, however a more general analysis of the precision of our scheme depending upon different post-selections have been performed in the following sections.

\section{Precision in measurement of temperature }
\label{sec4}

It is natural to wonder about the optimal temperature window where the present scheme works best, that is, most precisely. The usual procedure for determining the precision of quantum thermometers is through finding the corresponding quantum Cramer Rao bound. In this section, we adopt a different approach. We restrict to qubit systems for simplicity. However, the present analysis can be extended to higher dimensions in a similar fashion.

\subsection{Precision analysis for imperfect thermalization}

Let us assume that the initial pre-measurement state is not exactly a thermal state, but very close to it, and written in the following manner 

\begin{equation}
\rho_{T}^{(\delta)} = (1-\delta) \rho_T + \delta |\chi (\theta, \phi)\rangle\langle \chi (\theta, \phi)|,
\end{equation}
\noindent where $0 \leq \delta \ll 1$, and $|\chi \rangle$ is a random pure qubit state with corresponding Bloch sphere parameters $\theta$, and $\phi$. Physically this indicates imperfect thermalization, which is experimentally relevant, especially in situations where the thermalization timescale is not extremely fast compared to the time available for sensing. The corresponding weak value is denoted by $A_{w}^{\delta}$. Now, an experimentalist may use the formula \eqref{qubit_eqn} to find the apparent inverse temperature  $\tilde{\beta}$ of the bath as 

\begin{equation}
\tilde{\beta}= \frac{2}{E_2 - E_1} \arctanh (-i A_{w}^{(\delta)} )
\end{equation}

It is of obvious practical interest to us that the inferred value of temperature does not change wildly if the thermalization is imperfect. In order to quantify this, we invoke the idea of quantifying a relative error, which is the difference between the temperature furnished from imperfect thermalization, and the genuine temperature of the bath, divided by the bath temperature, and scaled by the imperfection $\delta$.  The squared relative error introduced through imperfect thermalization may  thus be taken to be  $|\tilde{\beta} - \beta|^2 / (\delta^2 \beta^2)$. It is this quantity we shall concentrate upon. To remove the effect of $|\chi\rangle$, which may conceivably capture some information about the state in which the probe was initialized, we shall finally average  over the pure states $|\chi\rangle$. We will write $E_2 - E_1$ as the gap $\Delta$ from here on. Performing a perturbation expansion for $A_{w}^{(\delta)}$ around $\delta = 0$,  and retaining terms upto first order in $\delta$, the expression for $\tilde{\beta}$ is given by

\begin{equation}
\tilde{\beta} = \frac{2}{\Delta} \arctanh \left( c_1 \tanh \frac{\beta \Delta}{2} - i c_2 \right),
\end{equation}
\noindent where $c_1 = 1 - \delta \frac{|\langle \psi_f| \chi\rangle|^2}{\langle \psi_f |\rho_T | \psi_f\rangle}$, and $c_2 = \delta \frac{\langle \psi_f|A|\chi\rangle \langle \chi|\psi_f\rangle}{\langle \psi_f|\rho_T|\psi_f\rangle}$. Note that, $c_1$ and $c_2$ are functions of the Bloch sphere angles $\lbrace \theta, \phi \rbrace$ of the state $\chi$. Thus, averaging over them, the resulting root mean squared relative error $\mathcal{N}_{\beta}$ for the particular weak measurement strategy adopted in the previous section, is written as a function of inverse temperature as 
\begin{widetext}
\begin{equation}
\mathcal{N}_\beta^{2} = \frac{1}{4 \pi} \iint \frac{|\tilde{\beta} - \beta|^2}{\beta^2 \delta^2} \sin\theta d\theta d\phi = \frac{1}{4\pi \delta^2} \int_{\theta = 0}^{\pi} \int_{\phi = 0}^{2 \pi} \left| \frac{2}{\beta \Delta} \arctanh \left[c_1 (\theta, \phi) \tanh \frac{\beta \Delta}{2}  - i c_2(\theta, \phi) \right] -1 \right|^2 \sin \theta d\theta d\phi .
\end{equation}
\label{defn_of_error}
\end{widetext}

\noindent Now, neglecting the second and the higher order coeffficients of $\delta$, we note that the expression for relative RMS error is written as 
\begin{equation}
\mathcal{N}_\beta^2 =  \frac{1}{4\pi} \iint \frac{4 |v_1 \tanh \frac{\beta \Delta}{2} + i v_2|^2}{\Delta^2 \beta^2 (1- \tanh^2 \frac{\beta \Delta}{2})^2}   \sin \theta d\theta d\phi 
\end{equation}

\noindent where $v_1 = (1 - c_1)/\delta$, and $v_2 = c_2/\delta$.  At this point, let us fix the observable $A =  -i |\psi_1\rangle\langle \psi_2| +i|\psi_2\rangle\langle \psi_1|$, as in the previous section, and assume the arbitrary post selected state $|\psi_f\rangle = \cos(\xi/2) |\psi_1\rangle + e^{i \nu}\sin (\xi/2) |\psi_2\rangle$. We also assume, without loss of generality, that the energy gap $\Delta  =1$. The expression for root mean squared relative error  $\mathcal{N}_{\beta}$ is given by
\begin{eqnarray}
 \frac{\sqrt{(13 - \cos 2\nu) \cosh \beta - 4 \cos 2\xi \sin^2 \nu \cosh^2 (\frac{\beta}{2}) - (3+ \cos 2\nu) }}{ 3 \beta \left[ \cosh (\beta/2) + \cos \xi \sinh(\beta/2) \right] [1 - \tanh^2(\beta/2)]}. \nonumber \\
 \label{all_postselected}
\end{eqnarray}
For specific choices of the post selected state, the anove equation yields the expression for error, and consequently, the inverse quantity signifies the precision of measurement. For example, assuming $|\psi_f\rangle =\frac{1}{\sqrt{2}} \left( |\psi_1\rangle + |\psi_2\rangle \right)$, as in the previous section, the relative error reads as 
\begin{equation}
\mathcal{N}_{\beta}^{+} = \sqrt{ \frac{1 + 4 \cosh \beta + 3 \cosh 2\beta}{9 \beta^2}}.
\end{equation}

\begin{figure}
\includegraphics[width = 0.3\textwidth]{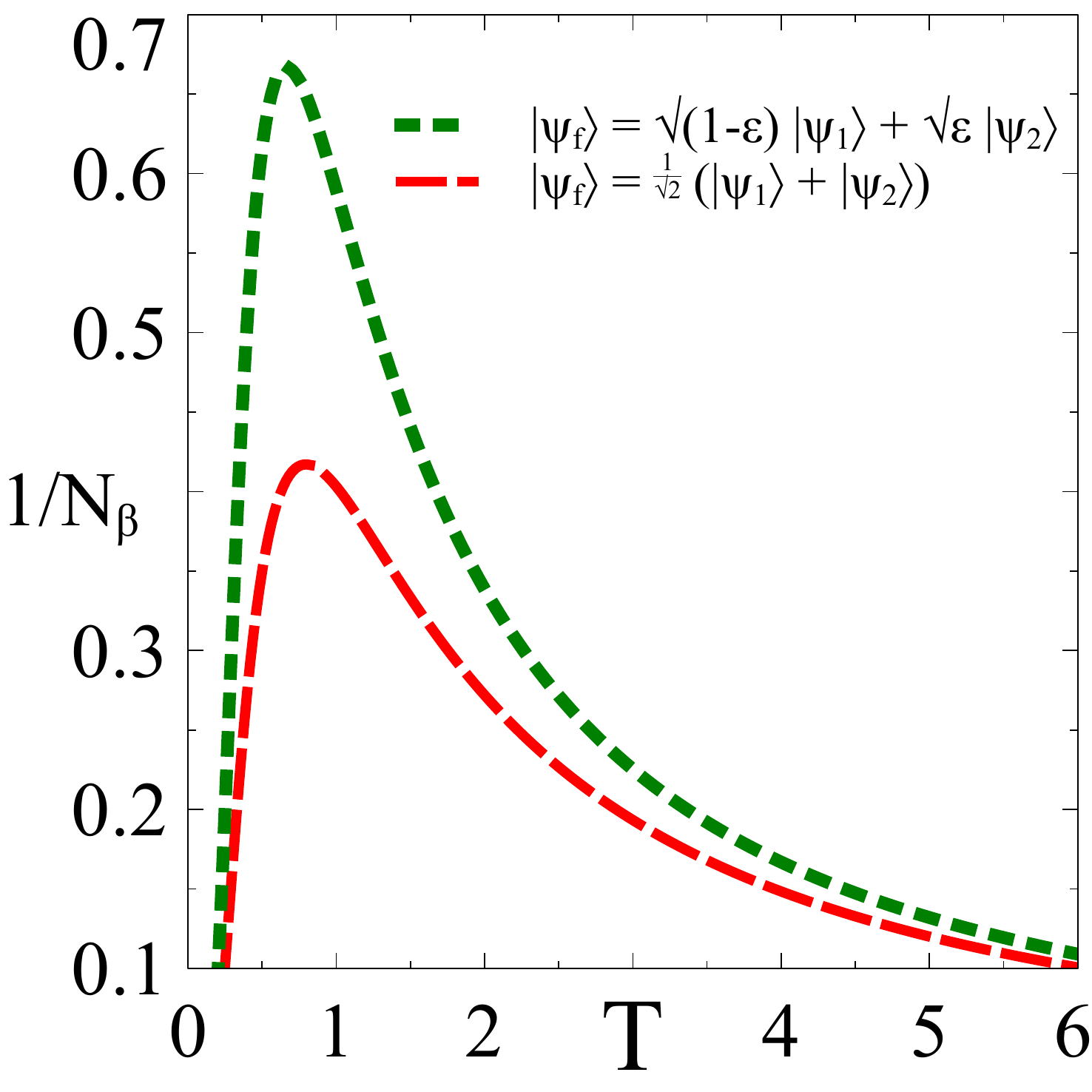}
\caption{ (color online) Precision of the scheme plotted against the temperature for the specific choice of post-selected states  $|\psi_f\rangle = \sqrt{1-\epsilon} |\psi_1 \rangle + \sqrt{\epsilon} |\psi_2\rangle$ (green dotted curve), $|\psi_f\rangle = \frac{1}{\sqrt{2}} \left( |\psi_1\rangle + |\psi_2\rangle \right)$ (red dashed curve). Energy gap between ground and excited states is unity in each case. $\epsilon = 0.01$ is assumed. }
\label{plot_precision_with_temp}
\end{figure}

\begin{figure}
\includegraphics[width = 0.45 \textwidth]{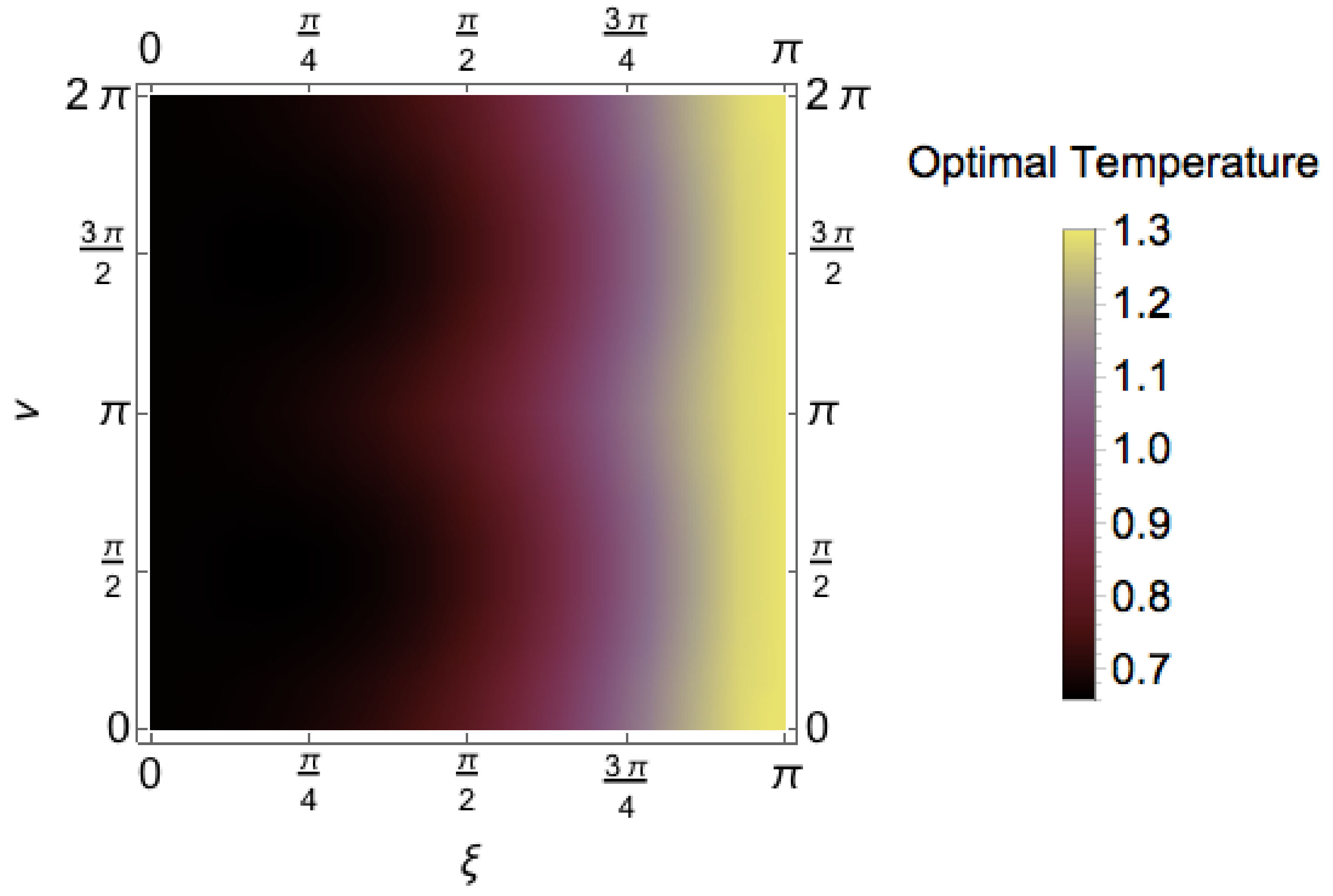}
\caption{(color online) The temperature at which the optimal precision is achieved is plotted against the parameters $\xi$ and $\nu$ of the post selected state $|\psi_f\rangle$.}
\label{plot_location}
\end{figure}

\begin{figure}
\includegraphics[width = 0.45\textwidth]{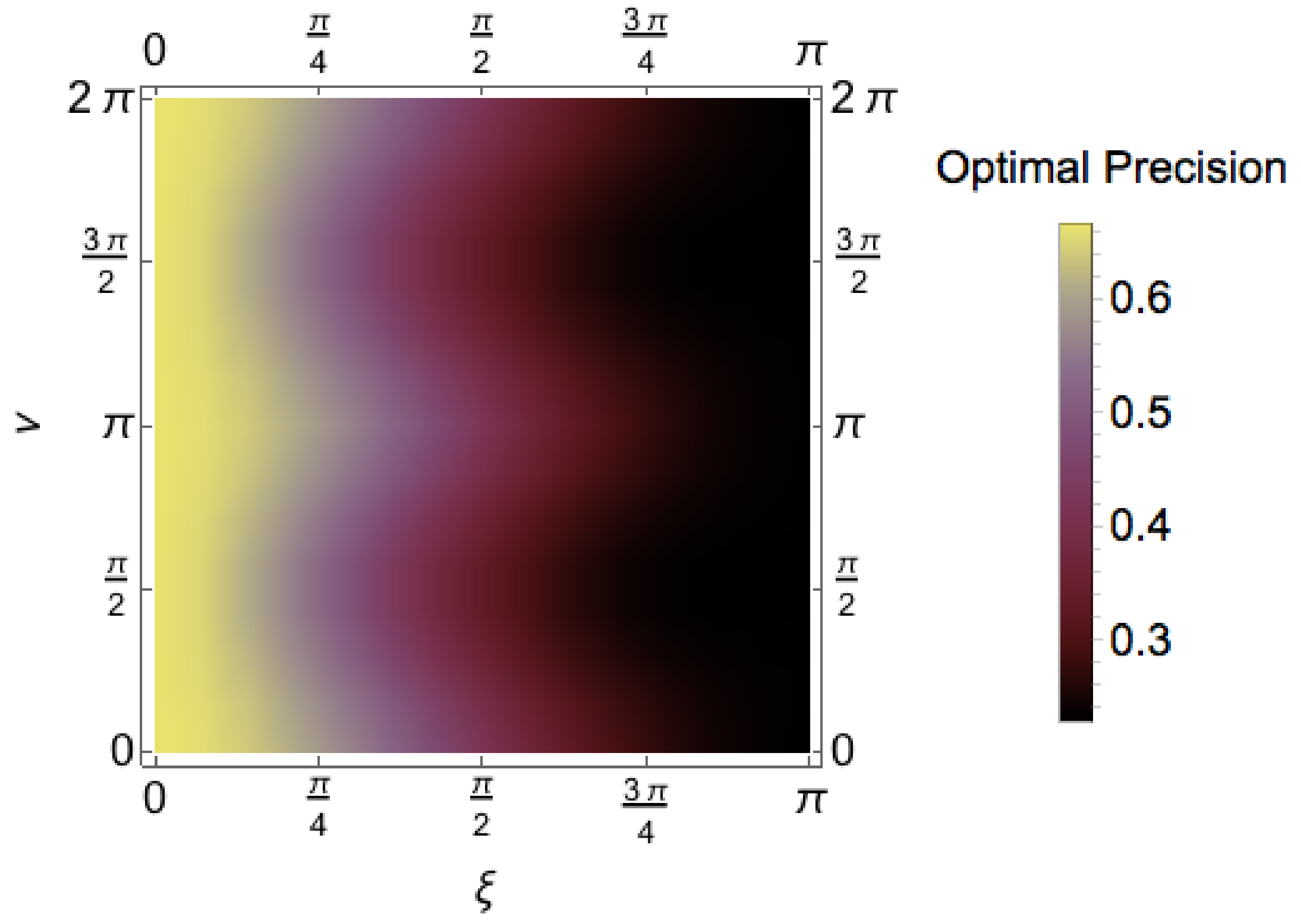}
\caption{ (color online) The magnitude of precision at optimal temperature is plotted against the parameters $\xi$ and $\nu$ of the post selected state $|\psi_f\rangle$. }
\label{plot_3d_value}
\end{figure}

\begin{figure}
\includegraphics[width = 0.45\textwidth]{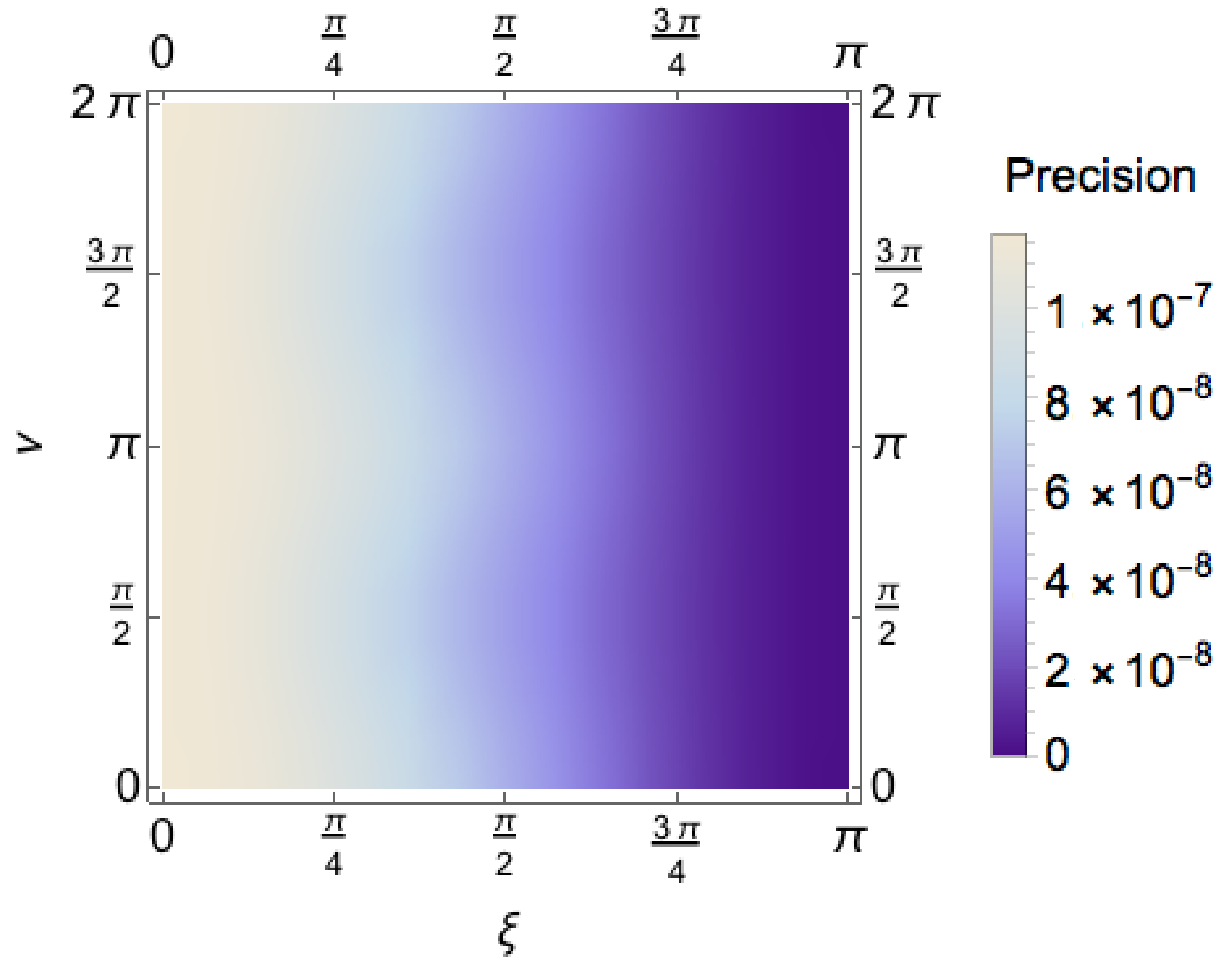}
\includegraphics[width = 0.45\textwidth]{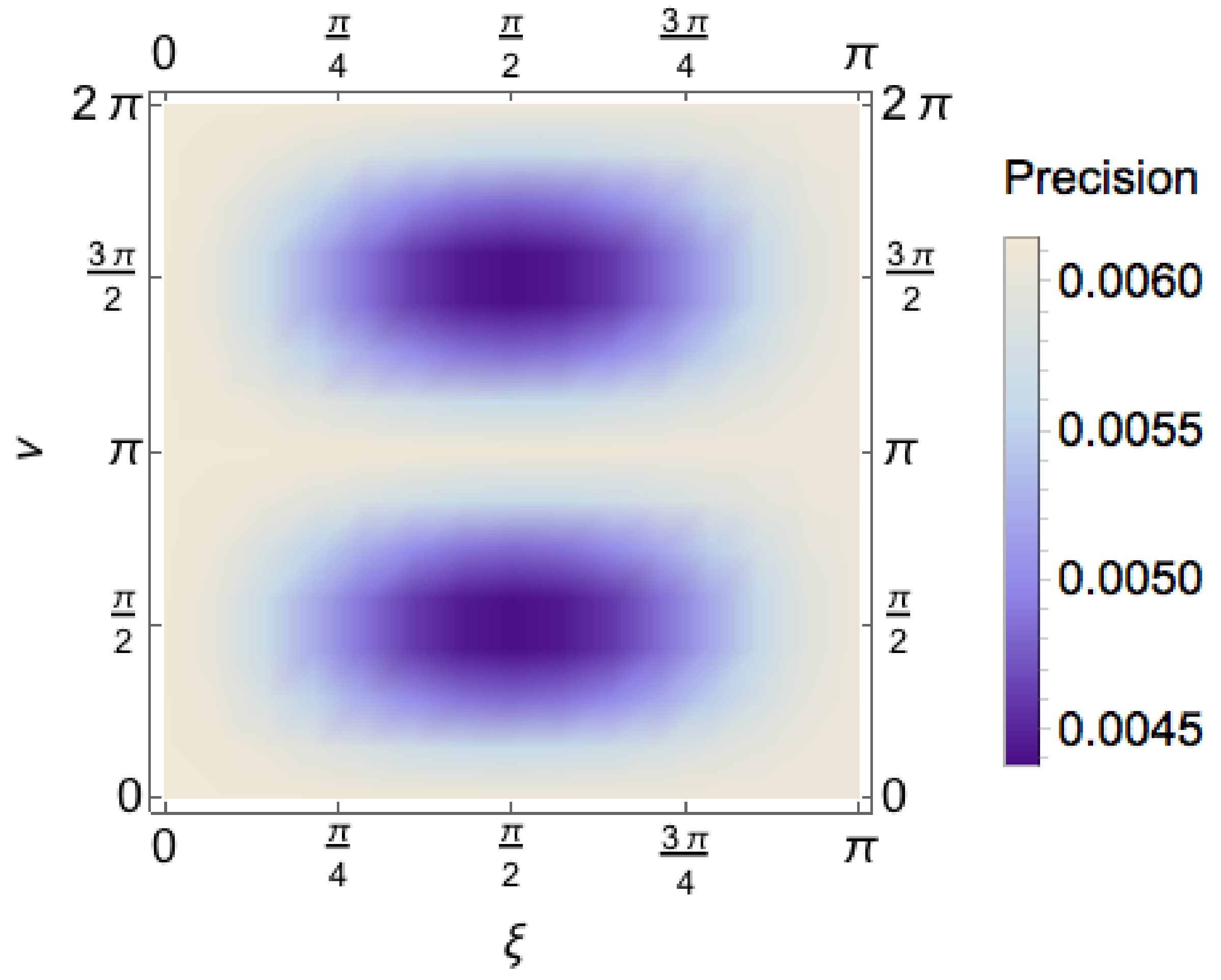}
\caption{(color online) The magnitude of precision is plotted against the parameters $\xi$ and $\nu$ of the post selected state $|\psi_f\rangle$ for (\emph{above}) low temperature T = 0.05, and (\emph{below}) high temperature T = 100.}
\label{plot_high_and_low}
\end{figure}

\noindent  Fig. \ref{plot_precision_with_temp} shows that the precision, which is defined as the inverse of the relative error, attains a relatively narrow peak at a finite temperature, which determines the best operating window of the scheme. While the qubit thermometric schemes \citep{stace, correa_prl, sanpera_review, our_thermometry} based on the strong measurement are different in conception than the present scheme, it is nonetheless noteworthy that the phenomenon of a narrow peak in the precision corresponding to an optimal temperature window is present in both the cases. We also note that for the QFI based analysis of optimal qubit thermometric probes with unit energy gap, the peak is situated at $T \approx 0.41$ \citep{stace, correa_prl}, which is obtained through solving the transcendental equation \citep{stace, correa_prl} $e^{1/T} = (1 + 2T)/(1 -2T) $. In comparison, in the present scheme, for a specific post-selected state $|\psi_f\rangle$, the location of the peak for optimal precision is obtained by the vanishing of the first derivative with respect to the inverse temperature $\beta$ of the expression in \eqref{all_postselected}. In particular, for the specific example $|\psi\rangle =  \frac{1}{\sqrt{2}} \left( |\psi_1\rangle + |\psi_2\rangle \right)$ discussed in the previous section, the equation takes the form 
\begin{equation}
\beta = \frac{3 \cosh \beta -1}{3 \sinh \beta - 2 \tanh \frac{\beta}{2}},
\end{equation} 
\noindent which has the solution $T \approx 0.79$.  The temperatures corresponding to optimal precision for other choices of post-selected states are depicted in Fig \ref{plot_location}. Interestingly, for every $|\psi_f\rangle$, the corresponding optimal temperature is significantly higher than $0.41$. This indicates the possibility that the present scheme may be better than the strong measurement based one based one for the relevant temperature range. It is natural to wonder about the post-selected state $|\psi_f \rangle$ which corresponds to maximum precision. From Fig \ref{plot_3d_value} as well as Fig \ref{plot_high_and_low}, it may be concluded that for any arbitrary temperature,  $|\psi_f\rangle$ close to $|\psi_1\rangle$ maximizes the precision. Here, let us add a note of caution, from the definition, the weak value is independent of the temperature when $|\psi_f\rangle$ is exactly $|\psi_1\rangle$, hence measuring the weak value furnishes no information about the temperature. Thus,  we must take $|\psi_f\rangle$ to be extremely close to $|\psi_1\rangle$, but not identically equal.

\subsection{Precision analysis for unsharp post-selection}

Let us now consider another potential source of error in the present scheme, that is, the post selection may be unsharp. Let us assume the unsharp post-measurement qubit state in the form
\begin{equation}
\rho_{f}^{(\epsilon)} = (1-\epsilon) |\psi_f\rangle\langle \psi_f| + \frac{\epsilon}{2} \openone
\end{equation}

\noindent Thus, the corresponding perturbed weak value of the observable $A$ is given by 
\begin{equation}
A_{w}^{\epsilon} = \frac{\text{Tr}(\rho_f^{(\epsilon)} A \rho_T)}{\text{Tr}(\rho_{f}^{(\epsilon)} \rho_T)} \approx A_{w}\left[1 + \frac{\epsilon}{A_w}\left(\frac{\text{Tr} (A \rho_T) - 1}{\langle \psi_f|\rho_T|\psi_f\rangle} \right) \right] + \mathcal{O}(\epsilon^2)
\end{equation}

\noindent Now, using \eqref{qubit_eqn}, the corresponding expression for shifted inverse  temperature is $\tilde{\beta} = \frac{2}{\Delta} \arctanh (-i A_{w}^{(\epsilon)})$. From which, the formula for the squared relative error incurred is given by 

\begin{equation}
\mathcal{N}_{\beta}^2 = \frac{|\tilde{\beta} - \beta|^2}{\epsilon^2 \beta^2} = \frac{4}{\Delta^2 \beta^2 [1 - \tanh^2 (\beta \Delta/2)]^2} \left|\frac{\text{Tr}(A \rho_T) -1}{\langle \psi_f|\rho_T|\psi_f\rangle} \right|^2.
\end{equation}

\noindent 
As in the previous section, if one chooses $A =  -i |\psi_1\rangle\langle \psi_2 | + i |\psi_2\rangle\langle \psi_1|$, then, assuming $\Delta =1$ without loss of generality, the corresponding expression for relative error reads as 
\begin{equation}
\mathcal{N}_{\beta} = \frac{4}{\beta \left[1 - \tanh^{2} \frac{\beta}{2}\right] (1 + \cos \xi \tanh \frac{\beta}{2})},
\end{equation}
\noindent where $\xi$ is the polar angle of the pure post selection state $|\psi_f\rangle$. For the specific choice $|\psi_f\rangle = \frac{1}{\sqrt{2}} \left( |\psi_1\rangle + |\psi_2\rangle \right)$ in the last section, this amounts to the following expression for precision, which is defined as the inverse of the error
\begin{equation}
1/\mathcal{N}_{\beta} (|+\rangle) = \frac{4}{\beta \left[1 - \tanh^{2} \frac{\beta}{2}\right]},
\end{equation}
\begin{figure}
\includegraphics[width = 0.45\textwidth]{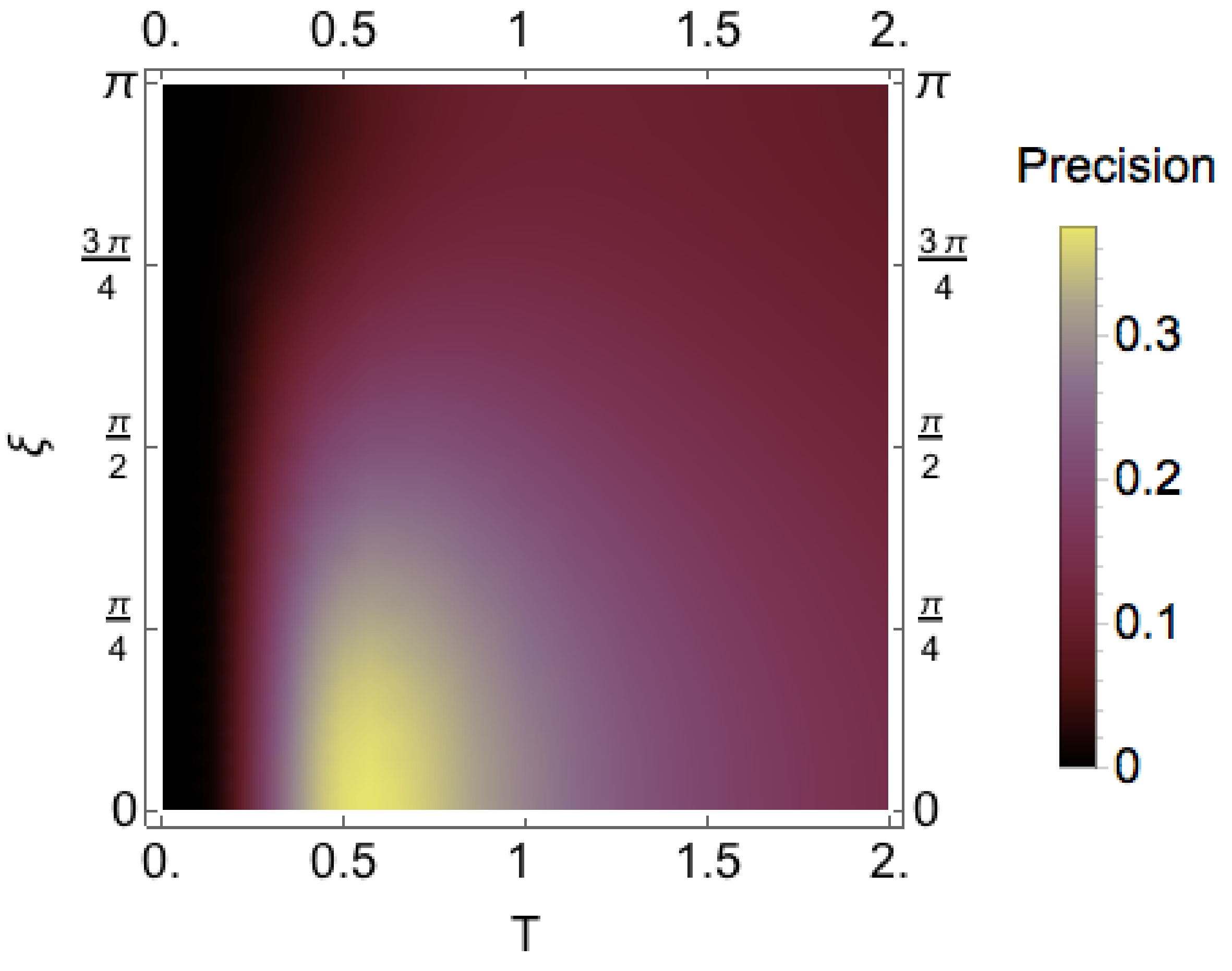}
\caption{ (color online) The magnitude of precision at optimal temperature is plotted against the temperature $T$ and the polar angle $\xi$ of the post selected state $|\psi_f\rangle$. } 
\label{fig_postsel_3d}
\end{figure}

\begin{figure}
\includegraphics[width = 0.22\textwidth]{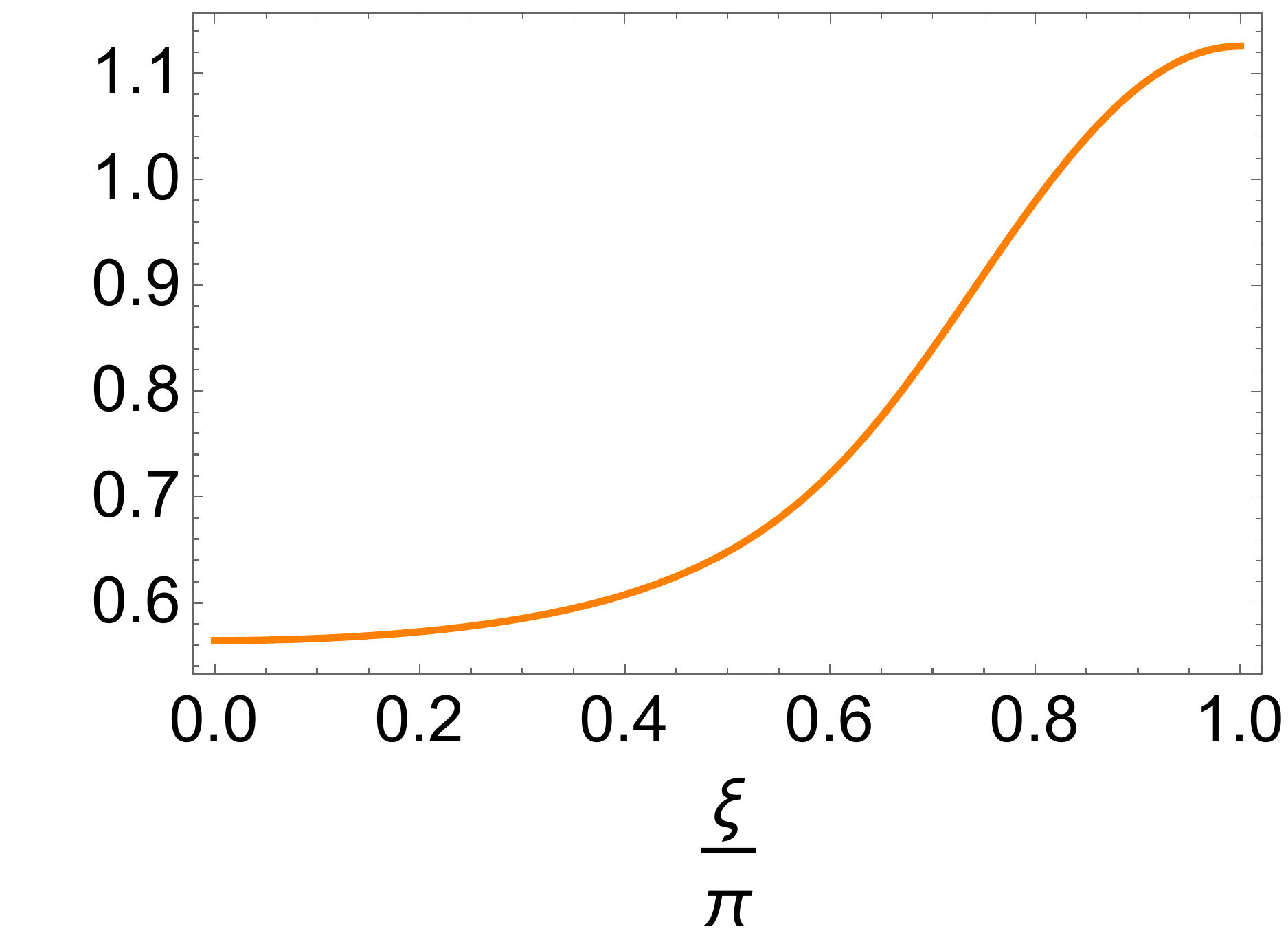}
\includegraphics[width = 0.22\textwidth]{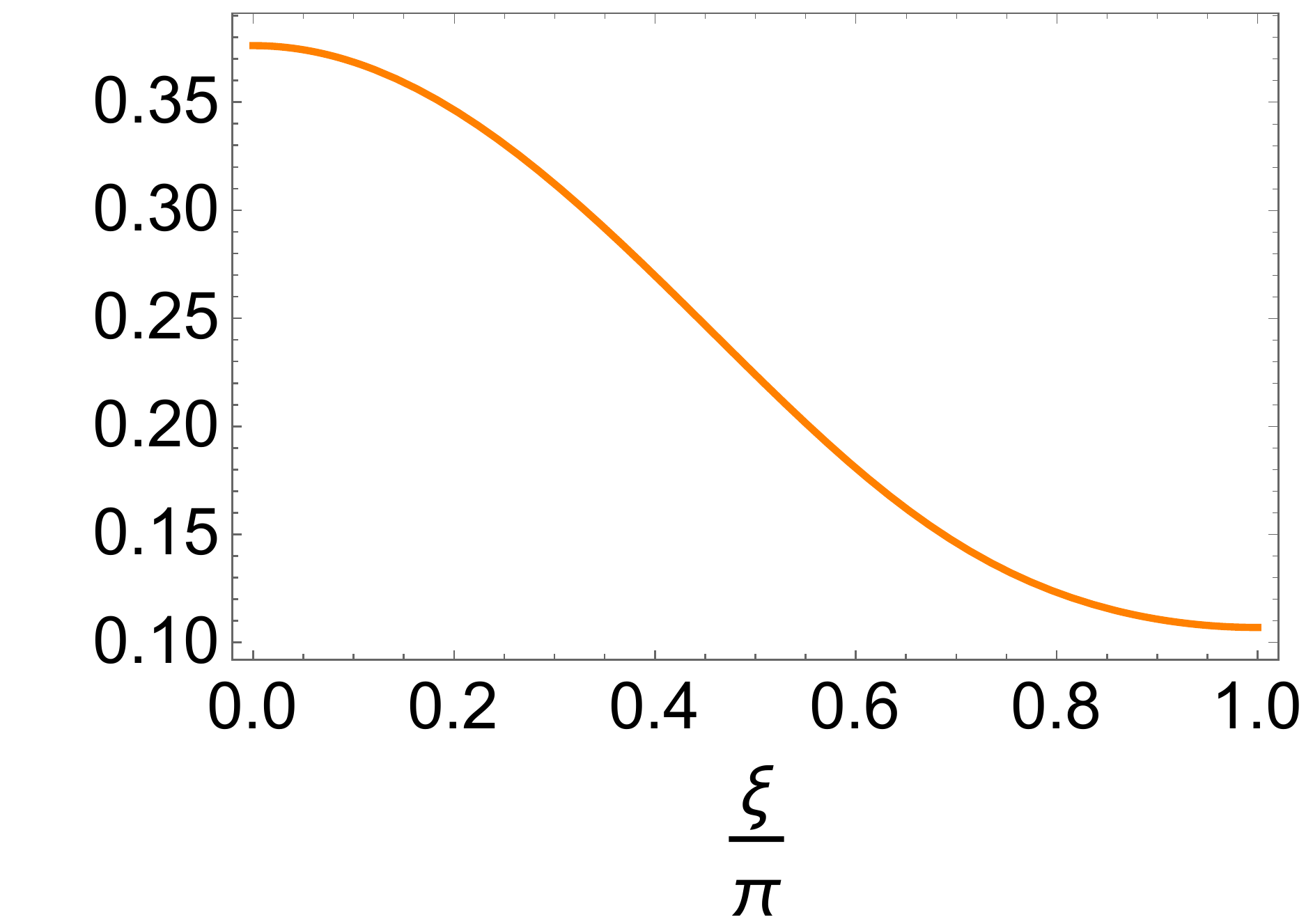}
\caption{ (color online) \emph{Left}- optimal temperature, and \emph{Right}- precision at optimal temperature, as a function of the polar angle $\xi$ of the post measurement state $|\psi_f\rangle$.}
\label{fig_postsel_2d}
\end{figure}

It  can be seen from Fig. \ref{fig_postsel_3d} that the precision, defined as the inverse of the relative error, attains a peak at some temperature, which determines the corresponding optimal temperature window for thermometry. Fig. \ref{fig_postsel_2d} reveals that the optimal temperature for the scheme varies from $T_{\text{opt}}\approx 0.54$ to $T_{\text{opt}} \approx 1.12$. Solution of the following transcendental equation determines the location of the optimal temperature $T^{*}$ for any given post-selected state $|\psi_f\rangle$ with the polar angle $\xi$
\begin{equation}
\cos \xi = \frac{\sinh \frac{1}{T^*} - T^* (1+ \cosh \frac{1}{T^*})}{T^* \sinh \frac{1}{T^*} - \cosh \frac{1}{T^*} +2}
\end{equation} 
\noindent An interesting feature  we observe in Fig \ref{fig_postsel_2d} is that there is a shift in the optimal temperature window towards the right, and higher temperatures is associated with the reduction in the optimal precision attainable through the present scheme. This is in line with the intuition that, as a distinctly quantum mechanical scheme, the weak measurement protocol should work best in the low temperature regime.

\section{Quantum Fisher Information based analysis of precision of weak thermometry protocol}

\label{sec5}
Until now, we have looked at the robustness of precision of the weak measurement based thermometric protocol. In this section, we present the complementary analysis of thermometric precision in this protocol through the usual quantum estimation theoretic methods. Let us first  recall the relevant bound on fluctuation $\Delta u$ of estimation of a single parameter $u$,  which is known as the \emph{quantum Cramer-Rao bound (QCRB)}.

\begin{equation}
\Delta u \geq \frac{1}{\sqrt{n \mathcal{F}_{u} (\rho_u)}},
\end{equation}

\noindent where $\mathcal{F}$ is the quantum Fisher Information for the state $\rho_{u}$, and $n$ is the number of runs. For the single parameter estimation case, it is always possible to saturate the lower bound. We now remember that the final state of the pointer after the post-selection in state $|\psi_f\rangle$ is given in \eqref{eqn2}, which is a pure state. For a pure state $|\psi_u\rangle$ with the corresponding parameter $u$, we recall that the QFI is given by \citep{petz}

\begin{equation}
\mathcal{F}_{u} = 4 \langle \dot{\psi_{u}} | \dot{\psi_{u}} \rangle - 4 | \langle \psi_u |\dot{\psi_u} \rangle |^2,
\label{qfi}
\end{equation}

\noindent where $|\dot{ \psi_u} \rangle$ denotes the first derivative of the state $|\psi_u \rangle$ with respect to the parameter $u$. Putting this in \eqref{eqn2}, we obtain the following expression for QFI for temperature $T$ after a little algebra 

\begin{equation}
\mathcal{F}_{T} = g^2 \tau^2 \left( \dfrac{dA_{w}}{dT} \right)^2 (\xi - \xi^2)
\end{equation}

\noindent Here $\xi = \langle \phi | e^{i g A_w^{*} \hat{P_{x}}} \hat{P_x}  e^{-i g A_w\hat{P_{x}}} |\phi \rangle = |\langle \phi|x\rangle|^2 e^{2g\tau \text{Im} (A_w)}$. Thus, upto leading order, the square-root of QFI is proportional to

\begin{equation}
\sqrt{\mathcal{F}_{T}} \propto \tilde{\mathcal{F}} =  \left| \dfrac{dA_{w}}{dT} \right|   = \left| \frac{\langle \psi_f | A \dfrac{d \rho_T}{dT}| \psi_f \rangle}{\langle \psi_f | \rho_T| \psi_f \rangle} - \frac{\langle \psi_f |  \dfrac{d \rho_T}{dT}| \psi_f \rangle \langle \psi_f | A \rho_T| \psi_f \rangle}{(\langle \psi_f | \rho_T| \psi_f \rangle)^2 }\right|
\end{equation}

Once more, we assume $\hat{A} = \hat{\sigma}_y$ unless otherwise mentioned. Expressing an arbitrary post-selected state as $|\psi_f\rangle = \cos \frac{\theta}{2} |0\rangle + \sin \frac{\theta}{2} e^{i\phi} |1\rangle$, we obtain the expression for scaled precision $\tilde{\mathcal{F}}$ as being 

\begin{figure}[h]
\includegraphics[width = 0.35\textwidth]{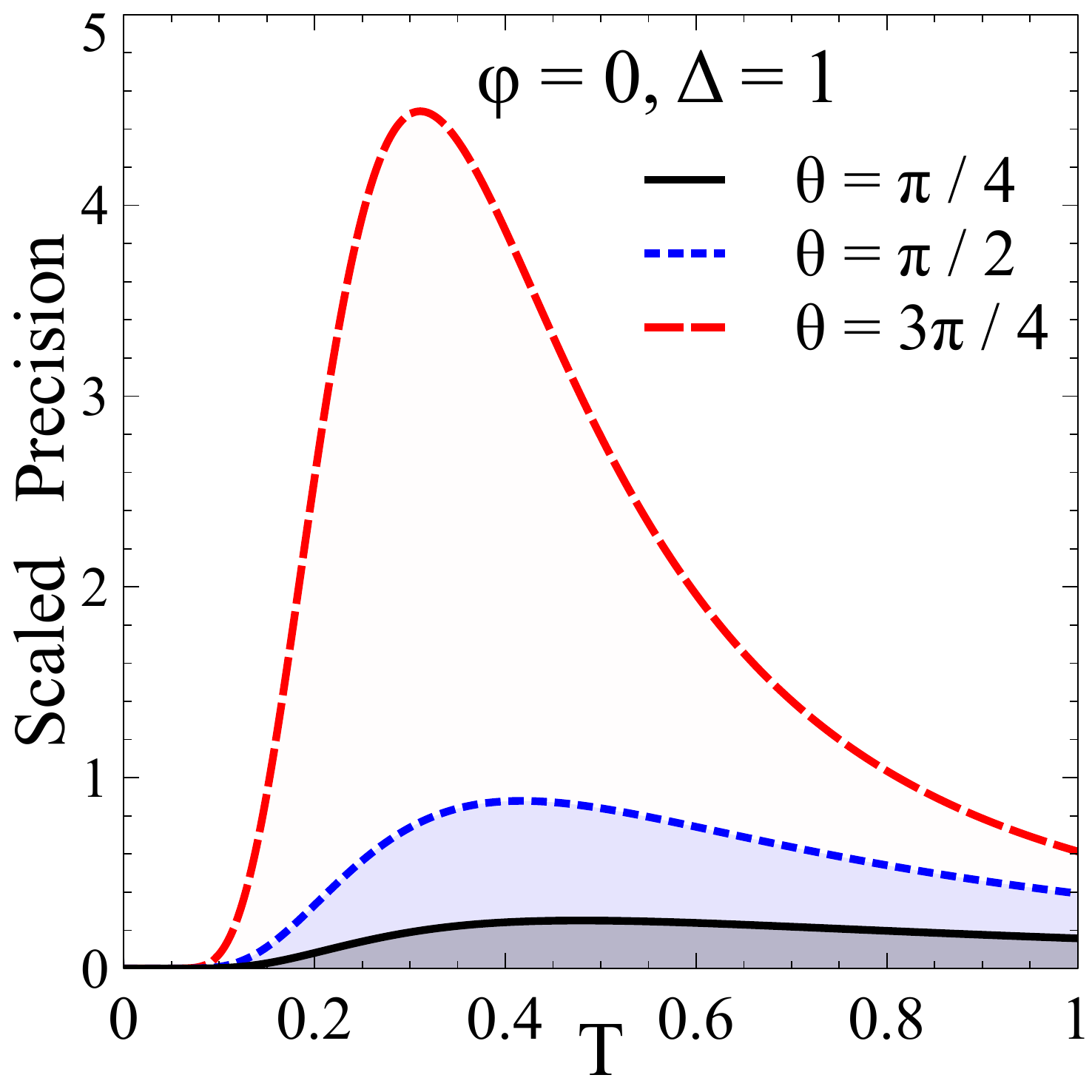}
\caption{ (color online) Scaled precision $\tilde{F}$ plotted against temperature for various choices of post-selected state parameter $\theta$. The energy gap $\Delta$ is taken to be unity, and the azimuthal angle $\phi$ of the post-selected state is assumed to be zero.}
\label{fig_fish}
\end{figure}

\begin{equation}
\tilde{\mathcal{F}} = \frac{2 e^{\Delta/T} \Delta \sin \theta \sqrt{\cos^2 \phi + \cos^2 \theta \sin^2 \phi}}{T^2 \left[1 - \cos \theta + e^{\Delta/T} (1+ \cos \theta) \right]^2}
\end{equation}

\noindent Some specific illustrations for different values of $\theta$ are demonstrated in Fig. \ref{fig_fish}. The existence of an optimal temperature window is once more observed, as is the feature of a shift in the optimal temperature window by shifting the choice of the post selection. As earlier, attempt to shift this optimal temperature window towards the right, i.e., to higher temperatures by judiciously choosing the post-selection state, inevitably results in a loss of optimal precision. Thus, the QFI based analysis yields the same qualitative picture as the analysis performed earlier.

\section{Discussions}
We have presented  a protocol for measuring temperature of a quantum mechanical bath through measuring weak values on a probe which is in thermal equilibrium with the bath. Our protocol relies on careful choice of Hamiltonian, which is generating the weak interaction, and the basis of post-selection measurement. Although our protocol is applicable to probes of any dimension, we restrict to qubit probes and compared our result with other thermometric schemes which employ strong POVM measurements on the probe and maximization of quantum Fisher information (QFI). We find that there is a narrow operational window of temperature values which can be optimally measured through our protocol, a feature characteristic to other usual  thermometric schemes. Moreover, in our protocol the peak value of the optimal temperature window is shifted towards higher temperature compared to the usual schemes. It may be helpful to investigate, in future works, the possible reason behind such a phenomenon. Interestingly, we find a trade off between shifting the optimal temperature window to higher temperatures and the optimal precision. 
%

We expect that the present work will turn out to be useful in the context of improving techniques for precise and efficient temperature estimation of nanoscale thermodynamic systems.  Several issues however, are left for future investigation. Firstly, a comprehensive description of the present scheme for higher dimensional probes, including harmonic oscillator probes, remains to be explored. In particular, we showed in the present work that the location of the optimal operating window depends on choice of post-selection, even for a probe with fixed energy spectrum. It is of some interest to determine whether the location of operating window can be made even more tunable by changing the post-selection in the higher dimensional cases.  Additionally, some other questions emerge from the idea of weak thermometry. Firstly, in the usual strong measurement based approach to thermometry, optimizing the precision for a qudit probe leads to the optimal energy spectrum of the probe to be a highly unphysical one, namely, a gapped ground state and all the other energy eigenstates are energetically degenerate \citep{correa_prl}. It is an open problem to figure out the form of the optimal energy spectrum of the probe in the arbitrary dimensional case, and check whether the corresponding Hamiltonian is experimentally feasible.. Additionally, finding the information disturbance tradeoff \citep{PhysRevA.97.032129} for thermometers working with weak measurement schemes is an interesting avenue of future work. Depending upon the choice of the actual physical systems for the probe, the operating windows for the precise measurement of temperature will have different signatures for systems under consideration, which we would like to explore in future. Finally, the weak value based thermometric scheme relies on having an identically prepared ensemble, and hence suffers from the drawback that single shot measurement results can not achieve arbitrary precision, and further improvement of thermometric precision with finite round of measurements based on the weak thermometry scheme should be useful to perform in future.

\section*{Acknowledgements}
CM acknowledges a doctoral fellowship from the Department of Atomic Energy, Government of India, as well as funding from INFOSYS. AKP and CM  wish to thank The Institute of Mathematical Sciences, Chennai, for their kind hospitality while hosting them for a visit, during which this work was completed. We would like to thank the anonymous referees for their constructive criticism.

\section*{Appendix}
In this appendix, we provide further details on calculations in sec. IV. Unless otherwise mentioned, equation numbers will correspond to equation numbers in the main text. 

\emph{Derivation of eqn. (21) - }  The integrand of eqn. (20) is given, according to the notation in the main text, as 

\begin{eqnarray}
&=&\left| \frac{2}{\beta \Delta} \arctanh \left[c_1 (\theta, \phi) \tanh \frac{\beta \Delta}{2}  - i c_2(\theta, \phi) \right] -1 \right|^2  \nonumber \\
&=&\left| \frac{2}{\beta \Delta} \arctanh \left[ (1- \delta v_1) \tanh \frac{\beta \Delta}{2}  - i \delta v_2 )\right] -1 \right|^2 \nonumber \\
&=& \left| \frac{2}{\beta \Delta} \arctanh \left[ \tanh \frac{\beta \Delta}{2}  - \delta \left(v_1 \tanh \frac{\beta \Delta}{2} + i v_2 \right)\right] -1 \right|^2 \nonumber \\
\end{eqnarray}

\noindent Expanding the last expression in terms of $\delta$ and retaining only upto first order yields the following expression for the integrand  

\begin{eqnarray}
&=& \left| \frac{2}{\beta \Delta} \left( \frac{\beta \Delta}{2} - \frac{\delta (v_1 \tanh \frac{\beta \Delta}{2} + i v_2}{1- \tanh^2 \frac{\beta \Delta}{2}}  \right) -1 \right|^2 \nonumber \\
&=&  \frac{4 \delta^2 | v_1 \tanh \frac{\beta \Delta}{2} + i v_2|^2}{\Delta^2 \beta^2 (1- \tanh^2 \frac{\beta \Delta}{2})^2},
\end{eqnarray}

which is the integrand in eqn. (21) of the main text. The integration on the next step is performed on Wolfram Mathematica, and the simplified result for our choice of observable $A$ is given out by the expression in eqn. (22).

\emph{Derivation of eqn. (23) from eqn. (22) --} From eqn. (22), for our specific choice of post-selection, the RMS error reads 

\begin{eqnarray}
\mathcal{N}_\beta &=& \sqrt{\frac{12 \cosh \beta - 4}{9 \beta^2 \cosh^2 \frac{\beta}{2} (1- \tanh^2 \frac{\beta}{2})^2}} \nonumber \\
&=& \sqrt{\frac{(12 \cosh \beta - 4) \cosh^2 \frac{\beta}{2}}{9 \beta^2}} \nonumber \\
&=& \sqrt{\frac{(6\cosh \beta - 2 + 6 \cosh^ \beta - 2 \cosh \beta}{9 \beta^2}} \nonumber \\
&=& \sqrt{\frac{3 (1+ \cosh 2\beta) + 4 \cosh \beta -2 }{9 \beta^2}} \nonumber \\
&=& \sqrt{\frac{1+ 4 \cosh \beta + 3 \cosh 2\beta}{9 \beta^2}}
\end{eqnarray}

\emph{Derivation of eqn. (24) -- } Optimizing the RMS error requires the derivative with respect to temperature $\partial_\beta \mathcal{N}_\beta$ to vanish. Hence, 
\begin{eqnarray}
 9 \beta^2 \left( 6 \sinh 2 \beta + 4 \sinh \beta \right) &=& 18 \beta \left( 1 + 3 \cosh 2\beta + 4 \cosh \beta \right); \nonumber \\
\Rrightarrow\beta &=& \frac{1+ 3 \cosh 2 \beta + 4 \cosh \beta}{3 \sinh 2 \beta + 2 \sinh \beta} ;\nonumber \\
\Rrightarrow\beta &=& \frac{4 \cosh^2 \frac{\beta}{2} \left(3 \cosh \beta -1\right)}{6 \sinh \beta \cosh \beta + 2 \sinh \beta} ;\nonumber \\
\Rrightarrow \beta &=&  \frac{3 \cosh \beta -1}{3 \sinh \beta - 2 \tanh \frac{\beta}{2}} 
\end{eqnarray}

\bibliographystyle{apsrev4-1}
\bibliography{temp_weak_value_project}

\end{document}